\documentstyle[12pt,epsf,amssymb]{article}

\begin{document}

\baselineskip=18.8pt plus 0.2pt minus 0.1pt
%\baselineskip=19pt plus 0.2pt minus 0.1pt

%%%%%% private definition %%%%%%%
\def\CR{\nonumber \\}
\def\pt{\partial}
\def\be{\begin{equation}}
\def\ee{\end{equation}}
\def\bea{\begin{eqnarray}}
\def\eea{\end{eqnarray}}
\def\eq#1{(\ref{#1})}
\def\la{\langle}
\def\ra{\rangle}
\def\hyp{\hbox{-}}
%%%%%%%%%%%%%%%%%%%%%%%%%%%%%%%%%

\begin{titlepage}
\title{
\hfill\parbox{4cm}
{\normalsize SU-ITP 01-02 \\ YITP-01-05 \\{\tt hep-th/0101208}}\\
\vspace{1cm}
Geometry on string lattice}
\author{
Naoki {\sc Sasakura}\thanks{\tt naokisa@stanford.edu,  
sasakura@yukawa.kyoto-u.ac.jp}
\\[7pt]
{\it Department of Physics, Stanford University,}\\
{\it Stanford, CA 94305-4060, USA}\\
and \\
{\it Yukawa Institute for Theoretical Physics, Kyoto University,}\\
{\it Kyoto 606-8502, Japan}\thanks{\tt Permanent address}}
\date{\normalsize January, 2001}
\maketitle
\thispagestyle{empty}

\begin{abstract}
\normalsize
Using the method developed by Callan and Thorlacius, 
we study the low energy effective geometry on a two-dimensional string 
lattice by examining the energy-momentum relations of the low energy 
propagation modes on the lattice.
We show that the geometry is identical for both the oscillation modes 
tangent and transverse to the network plane. 
We determine the relation between the geometry and the lattice variables.  
The lowest order effective field theory is given by the dimensional 
reduction of the ten-dimensional $N=1$ Maxwell theory.
The gauge symmetry is related to a property of 
a three-string junction but not of a higher order junction.
A half of the supersymmetries in the effective field theory should be broken
at high energy.
\end{abstract}
\end{titlepage}

\section{Introduction}
\label{Introduction}
\noindent
String junction \cite{junction1,junction2}
in type IIB string theory is known as a BPS
object preserving a quarter of the supersymmetries of type IIB string
theory \cite{BPS1,sen}. 
Connecting many of these junctions, string network
can be constructed \cite{sen}. When the whole network is stabilized within a 
two-dimensional plane, the network as a whole was also shown to be a quarter 
BPS state \cite{sen}. Protected by these remaining supersymmetries,
there exist rigid zero modes of string network \cite{zero1,zero2}. 
They are roughly the modes changing the sizes of each loop and can even 
reconnect the strings to change the topology of a network. 
Since these zero modes exist locally in a string network,
it is generally an object which can have a macroscopic entropy \cite{kol}.
In this paper, we will continue pursuing the possibility whether 
a string lattice can be regarded as a kind of continuous space 
in the low energy limit \cite{sasa}, like in lattice gauge theory.
This way of view might be similar in spirit to the idea of brane 
world \cite{brane,braneW} and also to the network appearing in quantum 
gravity \cite{Regge,Penrose,gravity1,gravity2,Ambjorn}.

A main question in regarding a string lattice as a continuous space is what
is the low energy dynamics on it.
The dynamics of a string network was analyzed in the small oscillation limit 
in \cite{Rey,Callan}, and a general framework for studying the moduli space
of the above mentioned zero modes was recently given in \cite{zerodyn}. 
The question of what kind of field theory is on a string lattice was 
studied in our previous paper \cite{sasa} for the case of a regular hexagonal
lattice. There the oscillations of a string lattice is approximated 
by the oscillations of junctions between which there are straight strings.
This approximation of straight strings would be valid 
when the motions of the string 
junctions are much slower than the motions of the oscillations on the strings.
Our main interest was that, since the zero modes exist locally in 
the network, they may appear as a local gauge symmetry in 
the low energy effective field theory.
In fact, we have obtained the result that the oscillation mode tangent 
to the network plane is described by Maxwell theory, 
the gauge symmetry of which comes from these zero modes.
As for the oscillations transverse to the network plane,
we have obtained a scalar field theory with seven scalars.

The low energy supersymmetries of the effective field theory give
a good relation between the Maxwell and the scalar modes.
From the supergravity argument of \cite{sen},
the eight supersymmetries remaining on the network 
are in the spinor representation {\bf 8} of the $SO(7)$
symmetry which is the rotation among the directions transverse to the network.
In \cite{Callan}, the eight supersymmetries were identified
from the world sheet view point.
There are one supersymmetry for each of the seven transverse directions and 
another for the mode along the network. 
Thus the remaining supersymmetries and the $SO(7)$ symmetry 
should mix up the scalar fields and the Maxwell field.

The approximate treatment in our previous paper \cite{sasa} can 
be well controlled by assigning a point-like mass to each junction. 
If we take the mass sufficiently large,
the oscillations of the junctions will become sufficiently slower than
the string oscillations, which will validate our approximation.
However, in the limit of vanishing mass, which is the case we are interested 
in, we cannot fully believe the validity of our approximation.

In this paper we will instead use the method developed in \cite{Callan}
to investigate the low energy dynamics of a periodic string lattice, and
will generalize our previous results \cite{sasa}.
In section \ref{gaugesym}, we will recapitulate and discuss some properties 
of the scattering matrix associated to junctions and the low energy 
oscillation modes on a string lattice. 
We will show that the gauge symmetry is a general feature of a string lattice
made of three-string junctions, but not of one made of higher order junctions.
In section \ref{geometry},
we will extract the low energy geometry by studying the energy-momentum
relation of the low energy propagation modes.
The geometry is identical for both the tangent and transverse modes,
so that one geometry can be assigned to a string lattice.
It is also shown to be consistent with the gauge symmetry. 
In section \ref{super},
we will discuss the low energy effective field theory and 
its supersymmetry. 
Section \ref{sum} is devoted for summary and discussions.

\section{The gauge symmetry}
\label{gaugesym}
In this paper we will use the approach to the dynamics of 
the small oscillations of string junction which was developed in
\cite{Callan}.
Their approach is to relate the in-wave and out-wave at a string
junction by a matrix to characterize the dynamics of the string 
junction. In their small oscillation limit, the dynamics is independent
for the mode in each of the directions transverse to the junction plane 
(out-plane) and the mode along the junction plane (in-plane). 
Let us consider the fluctuations $\phi_i(x_i,t)$ on a string
in one of those independent directions, where $i\ (i=1,\cdots,N)$ 
is the index for the strings forming an $N$-order string junction. 
The oscillation on a string can be written as
\be
\label{phi}
\phi_i(x_i,t)=Re\{ (A_i \exp(i \omega x_i) + B_i \exp(-i\omega x_i))
\exp(-i\omega t)\},
\ee
where $A_i,B_i$ are the complex mode amplitudes of out-wave and 
in-wave, respectively,
and $x_i>0$ is the distance from the junction measured along the $i$-th string.
The oscillations slightly deform the configuration of the strings 
near the string junction so that they satisfy the following two physical 
matching conditions at the junction.
One is the continuity condition that the strings should meet at one 
point, and the other is the tension balance condition that 
the forces from the string tensions pulling the junction 
should add up to vanish.
From these conditions they determined the matrix relating the in-wave and 
out-wave amplitudes, $\vec{A}=S \vec{B}$, for a general $N$-order string 
junction. In the case of a three-string junction ($N=3$), 
the matrix $S$ is given by
\be
S_{out\hyp plane}=-{\bf 1} + \frac2{\sum_{i=1}^3 t_i}\left( 
\begin{array}{ccc}
t_1 & t_2 & t_3 \cr
t_1 & t_2 & t_3 \cr
t_1 & t_2 & t_3 
\end{array}
\right)
\ee
for an out-plane scattering, and, for an in-plane scattering,  
\be
S_{in\hyp plane}={\bf 1} -\frac2{D} \left( 
\begin{array}{ccc}
t_2 t_3 \sin^2 \theta_{23} & t_2 t_3 \sin \theta_{23} \sin \theta_{31}& 
t_2 t_3 \sin \theta_{23} \sin \theta_{12} \cr
t_1 t_3 \sin \theta_{31} \sin \theta_{23} & t_1 t_3 \sin^2 \theta_{31} & 
t_1 t_3 \sin \theta_{31} \sin \theta_{12} \cr
t_1 t_2 \sin \theta_{12} \sin \theta_{23} & t_1 t_2 \sin \theta_{12} 
\sin \theta_{31}& t_1 t_2 \sin^2 \theta_{12} 
\end{array}
\right),
\ee
where $t_i$'s and $\theta_{ij}$'s are the string tensions and the angles 
between the string $i$ and $j$, respectively, and 
$D=t_1 t_2 \sin^2 \theta_{12} + t_2 t_3 \sin^2 \theta_{23}
+t_3 t_1 \sin^2 \theta_{31}$.

A fairly convenient coordinate can be taken for these 
matrices \cite{Callan}. 
After a similarity transformation, $\hat S = \sqrt{t} S \sqrt{t}^{-1}$,
where $t$ is a diagonal matrix whose entries are the tensions,
they found
\be
\hat S_{out\hyp plane} = -{\bf 1} + 2 \vec{y} \otimes \vec{y},
\ee
where $\vec{y}$ is a unit vector
\be
\vec{y} =\frac1{\sqrt{\sum t_i}} (\sqrt{t_1},\sqrt{t_2},\sqrt{t_3}),
\ee
and
\be
\hat S_{in\hyp plane} = {\bf 1}- 2 \vec{z} \otimes \vec{z},
\ee
where $\vec{z}$ is a unit vector
\be
\vec{z}=\frac1{\sqrt{D}} (\sqrt{\tau_{23}},\sqrt{\tau_{31}},\sqrt{\tau_{12}})
\ee
with $\tau_{ij}=t_i t_j \sin^2\theta_{ij}$. 
The eigenvalues of these matrices are $(1,-1,-1)$ and $(1,1,-1)$, 
respectively, and the eigenvalues 1 and -1 correspond to Neumann and
Dirichlet boundary conditions generalized to a junction formed by
strings with arbitrary tensions, respectively.

Although they did not mention explicitly in their paper, 
the vectors $\vec{y}$ and $\vec{z}$ are the same.
This can be shown from the relations among the angles $\theta_{ij}$ and 
the tensions $t_i$. Due to the tension balance condition, the angles 
$\pi-\theta_{ij}$ and the tensions $t_i$ are in fact the angles and 
the edge lengths of a closed triangle, respectively, and so they 
should satisfy 
\be
\label{triangle}
\frac{t_1}{\sin \theta_{23}}=\frac{t_2}{\sin \theta_{31}}
=\frac{t_3}{\sin \theta_{12}}.
\ee
From these equations we obtain
\be
(\tau_{23},\tau_{31},\tau_{12})= C (t_1,t_2,t_3),
\ee
where $C$ is an unimportant positive factor.  Thus we obtain
\be
\vec{y}=\vec{z}.
\ee
This gives a simple relation between the out-plane and in-plane 
scattering matrices:
\be
\label{souteqsin}
S_{out\hyp plane}=-S_{in\hyp plane}.
\ee
As we will see, because of this relation, the spectra of the oscillation 
modes on a periodic string lattice in the tangent and transverse directions
are basically the same, and 
this is consistent with the discussions about the remaining 
supersymmetries in Sec.\ref{Introduction}.

As for a higher order string junction, the eigenvalues of the scattering 
matrices are $(1,-1,\cdots,-1)$ and $(1,1,-1,\cdots,-1)$ for out-plane and 
in-plane, respectively \cite{Callan}. 
Hence we can not expect a simple relation like \eq{souteqsin} to hold 
between the two kinds of scattering matrices in this case.
Thus, when a string lattice is made of higher order junctions, 
it is obscure how the low energy field theory respects the 
eight remaining supersymmetries expected from the supergravity argument.

The spectra of the oscillations on an infinite periodic string lattice 
can be analyzed by using the scattering matrices.
The case that a string lattice is given by a periodic connection of one 
kind of three-string junction was analyzed in \cite{Callan}.
The basic equations are given by
\bea
\label{basicrel}
\vec{A}&=&S\ \vec{B}, \cr
P\ \alpha\ \vec{B}&=&S\ P^*\ \alpha\ \vec{A}
\eea
where $S$ is $S_{out\hyp plane}$ or $S_{in\hyp plane}$ for the out-plane
or in-plane modes, respectively.
Here 
$P={\rm diag}(\exp(-i \omega l_1),\exp(-i \omega l_2),\exp(-i \omega l_3))$ 
is a diagonal matrix accounting for the phase shifts at the 
other ends of each string with length $l_i$ forming a three-string junction,
and $\alpha={\rm diag}(\exp(i \alpha_1),\exp(i \alpha_2),\exp(i\alpha_3))$ 
is a diagonal matrix of the phase shifts associated to the discrete 
translation symmetries of the periodic lattice. 
The momenta of the oscillation modes are related to 
$\alpha$ by
\be
\label{defalpha}
\alpha_i=\vec{l}_i\cdot (p_1,p_2),
\ee
where $\vec{l}_i$ denote the 
vectors in the two-dimensional lattice plane with the directions and 
lengths of the strings starting from the junction point, and $\cdot$
is the inner product between vectors on the plane.
The mode spectra, which give relations between $\omega$ and $(p_1,p_2)$, 
can be determined by analyzing the condition that the equations 
\eq{basicrel} have nontrivial solutions for $\vec{A},\vec{B}$.
Because of \eq{souteqsin}, the problem is basically equivalent
for the out-plane and in-plane cases.

As was noticed in \cite{Callan}, there is a special solution to 
the equations \eq{basicrel}. This solution exists for any 
momenta $(p_1,p_2)$ without any cost of energy: $\omega=0$.
These modes were identified as the gauge symmetry of 
the low energy effective action in our previous paper \cite{sasa}.
The reason for this identification is that this fact shows that these modes
are the time-independent local deformations of the string lattice
configuration under which the potential energy remains unchanged.
Let us look into the modes in more detail by solving \eq{basicrel} for
$P={\bf 1}\ (\omega=0)$. Since $\alpha$ is 
commutative with the tension matrix $t$, the problem of solving
\eq{basicrel} is easier after the similarity transformation by 
$\sqrt{t}$. Then, by eliminating $\vec{B}$, we obtain
\be
\label{eqofa}
\vec{\hat{A}} = \hat S_{in\hyp plane} \hat S_\alpha  \vec{\hat{A}},
\ee
where 
\be
\hat S_\alpha= 1- 2 (\alpha \vec{y})^* \otimes (\alpha \vec{y}).
\ee
Thus the solution to \eq{eqofa} for $\vec{\hat{A}}$ is given explicitly
by a vector orthogonal both to $\vec{y}$ and $\alpha \vec{y}$:
\be
\label{gaugesymA}
\vec{\hat{A}}_{gauge}=(\alpha \vec{y}) \times \vec{y},
\ee
where $\times$ denotes an external product in the three-dimensional
string index space.
As we mentioned, the equations for the oscillation spectra 
are basically equivalent for the out-plane and in-plane,
but the physical spectra are different. From \eq{phi}, for $\omega=0$,
the physical quantities are 
$(\vec{A}+\vec{B})_{out\hyp plane}=(1-S_{in\hyp plane})\vec{A}$ 
and $(\vec{A}+\vec{B})_{in\hyp plane}=(1+S_{in\hyp plane})\vec{A}$ 
for the out-plane and in-plane, respectively. 
Since the solution satisfies
$\hat S_{in\hyp plane}\vec{\hat{A}}_{gauge}=\vec{\hat{A}}_{gauge}$, the gauge
mode $\vec{\hat{A}}_{gauge}$ does not exist as an out-plane mode, while
it does as an in-plane mode. This makes a difference between the 
low energy effective field theories for the oscillations transverse and 
tangent to the lattice plane.

As for a higher order junction ($N\geq4$), the existence of the gauge mode is 
not expected. This can be shown as follows. From the analysis of
\cite{Callan}, the rescaled matrix of the in-plane scattering matrix 
for an $N$-string junction, 
${\hat S}_{in\hyp plane}=\sqrt{t}S_{in\hyp plane}\sqrt{t}^{-1}$, 
should have the from
\be
{\hat S}_{in\hyp plane}={\bf 1}-2 \sum_{i=1}^{N-2} \vec{z}_i \otimes \vec{z}_i,
\ee
where $\vec{z}_i$'s are orthogonal unit vectors. 
By a similar argument as above, in this $N$-junction case, 
\be
{\hat S}_\alpha={\bf 1}-2 \sum_{i=1}^{N-2} (\alpha \vec{z}_i)^* \otimes 
(\alpha \vec{z}_i).
\ee
Since there does not seem to exist any necessity for the vectors
$\{\vec{z}_i, \alpha \vec{z}_i\}$ to degenerate to less than $N$-dimensions, 
we do not expect that there exists a vector orthogonal to all 
$\{\vec{z}_i, \alpha \vec{z}_i\}$ for $N \geq 4$. 
This fact can be explicitly checked for some simple examples.
This is consistent with the argument that the gauge symmetry comes from
the zero modes of changing the sizes of each loop of
a network \cite{sasa}. 
We cannot change the size of a loop without changing the sizes of 
the other neighboring loops when higher-order junctions
are forming the loop. Thus we do not obtain local zero modes in such
a case. 

\section{Geometry on string lattice}
\label{geometry}
As for the interest in the possibility to regard a  
string lattice as a continuous space manifold in the low energy limit,
it would be interesting to investigate the effective low energy geometry 
on a string lattice.
This will be extracted from the relations between the energy and momentum
of the low-energy propagation modes, $\omega^2=\sum_{m,n=1}^2 g^{mn}p_mp_n$.

The condition for the equations \eq{basicrel} to have non-trivial
solutions was obtained in \cite{Callan}:
\bea
\label{eqCT}
0&=&(\sum_{i=1}^3 t_i^2)s_1s_2s_3+2t_1t_2s_3(\cos(\alpha_1-\alpha_2)-c_1c_2)
+2t_2t_3s_1(\cos(\alpha_2-\alpha_3)-c_2c_3) \cr
&&\ \ \ \ \ \ \ \ \ \ \ \ \ \ \ +2t_3t_1s_2(\cos(\alpha_3-\alpha_1)-c_3c_1),
\eea
where $s_i=\sin(\omega l_i),c_i=\cos(\omega l_i)$.
As we have discussed in the previous section, there is a rather 
trivial solution with $\omega=0$ for any given $(p_1,p_2)$. 
In addition, there exists another low energy 
solution in \eq{eqCT}. This is the low energy propagation mode
which we are interested in in this section.
Expanding \eq{eqCT} in the cubic orders of $\omega$ and the 
quadratic orders of $\alpha_i$, 
we obtain
\be
\omega^2=\frac1L 
(t_1 t_2 l_3 (\alpha_1-\alpha_2)^2+t_2 t_3 l_1 (\alpha_2-\alpha_3)^2
+t_3 t_1 l_2 (\alpha_3-\alpha_1)^2),
\ee
where
\be
L=(t_1l_2l_3+t_2l_3l_1+t_3l_1l_2)(l_1t_1+l_2t_2+l_3t_3).
\ee
Thus, substituting this with \eq{defalpha}, 
the two-dimensional metric tensor is obtained as 
\be
\label{metric}
g^{mn}=\frac1L\sum_{i<j,k\neq i,j}^3 
t_it_jl_k(\vec{l}_i-\vec{l}_j)^m(\vec{l}_i-\vec{l}_j)^n.
\ee

A more convenient expression of the metric tensor can be obtained as 
follows.
Let us pick up the terms with the form $\vec{l}_1\otimes\cdots$ in 
the numerator of the metric tensor \eq{metric}:
\be
\label{l1terms}
\vec{l}_1\otimes (t_1t_2l_3(\vec{l}_1-\vec{l}_2)
-t_3t_1l_2(\vec{l}_3-\vec{l}_1)).
\ee
Let us define the tension vectors in the two-dimensional lattice plane
by $\vec{t}_i=(t_i/l_i)\vec{l}_i$ (no sum). From the tension balance
condition, these vectors satisfy 
\be
\label{tensionbal}
\sum_{i=1}^3 \vec{t}_i=0.
\ee
Substituting $\vec{l}_i$ with $\vec{t}_i$ in \eq{l1terms}
and using \eq{tensionbal}, we obtain 
\be
(t_1l_2l_3+t_2l_3l_1+t_3l_1l_2)\ \vec{l}_1\otimes \vec{t}_1.
\ee
Thus another equivalent expression for the metric tensor is given by
\be
\label{metricanot}
g^{mn}=\frac{\sum_{i=1}^3 (\vec{l}_i)^m (\vec{t}_i)^n}
{\sum_{i=1}^3 l_i t_i}.
\ee
This expression clearly shows that, for any momentum $(p_1,p_2)$, 
the inequality $\omega^2=g^{mn}p_mp_n<\sum_{n=1}^2 p_n p_n$ holds.
This means that the velocity of the low energy propagation mode is always 
smaller than the light velocity for an observer in the target space.

Now let us study the polarization of the gauge symmetry obtained
in the previous section. 
Since the definition of the momentum \eq{defalpha} is through the inner 
product with variables in the target space, the momentum has a contravariant 
index, while the polarization of the gauge symmetry has a covariant index 
since the field $\phi$ of \eq{phi} denotes string motions in the target
space. Thus we need the metric tensor to see the relation between 
the momentum and the polarization.  
We will show below that the polarization is 
certainly parallel to the momentum when we take into account 
the metric \eq{metricanot} (or \eq{metric}).

From the result \eq{gaugesymA}, in the first order of the 
momentum, the in-wave amplitude of the gauge symmetry is given by
\bea
\label{inwavegauge}
(A_1,A_2,A_3)_{gauge}&=&\sqrt{t}^{-1} 
(\hat{A}_1,\hat{A}_2,\hat{A}_3)_{gauge} \cr
&=& \left( \sqrt{t_2t_3/t_1}(\vec{l}_2-\vec{l}_3)\cdot \vec{p},
\sqrt{t_3t_1/t_2}(\vec{l}_3-\vec{l}_1)\cdot \vec{p},
\sqrt{t_1t_2/t_3}(\vec{l}_1-\vec{l}_2)\cdot \vec{p}\right)
\eea
up to an unimportant all over factor.
Since $\vec{A}+\vec{B}=2\vec{A}$ in this case, 
\eq{inwavegauge} represents the physical polarization.
To extract the two-dimensional in-plane polarization vector from the data 
\eq{inwavegauge}, we will proceed as follows. 
Let us denote the corresponding two-dimensional polarization vector 
by $\vec{\varphi}_{gauge}$, and suppose it is expanded by the tension
vectors: $\vec{\varphi}_{gauge}=\sum_{i=1}^3 \varphi_i \vec{t}_i$.\footnote{
Since $\sum_{i=1}^3 \vec{t}_i=0$, this is a degenerate expression. In general
this expansion will not work well, but for our present purpose, 
this coordinate will turn out to be very convenient.}
Then, since $A_i$ denote the fluctuations transverse to each string $i$,
we have
\bea
A_i&=&\vec{\varphi}_{gauge}\cdot R(\vec{t}_i)/t_i \cr
&=&\sum_{j=1}^3 \varphi_j \vec{t}_j \cdot R(\vec{t}_i)/t_i \cr
&=&\sum_{j=1}^3 \varphi_j t_j \sin(\theta_{ij}),
\eea
where $R(\cdot)$ denotes a two-dimensional rotation operator of degree $\pi/2$.
Thus, using \eq{triangle} and setting $T=t_1/\sin(\theta_{23})=\cdots$,
we obtain
\be
(t_1A_1,t_2A_2,t_3A_3)=\frac{t_1t_2t_3}T\ (1,1,1)\times 
(\varphi_1,\varphi_2,\varphi_3).
\ee
Comparing with the data \eq{inwavegauge}, we obtain
\be
(\varphi_1,\varphi_2,\varphi_3)=\frac{T}{\sqrt{t_1t_2t_3}}
(l_1\cdot p,\ l_2\cdot p,\ l_3\cdot p).
\ee
Thus we finally obtain the desired result
\bea
(\vec{\varphi}_{gauge})^m&=&\frac{T}{\sqrt{t_1t_2t_3}} \sum_{i=1}^3
(\vec{t}_i)^m (\vec{l}_i\cdot p) \cr
&=& C g^{mn}p_{n},
\eea
where $C$ is an unimportant factor.

\section{Effective field theory and supersymmetry}
\label{super}
Let us first discuss the effective action for one of the oscillations 
transverse to the network plane.
In the small oscillation limit, the kinetic term of the effective action 
should be well approximated by 
\be
\label{kinact1}
K=\int dt\ \frac{m}2 \sum_{j=junction} \left(\frac{d\phi_{j}}{dt}\right)^2,  
\ee
where $m=(l_1t_1+l_2t_2+l_3t_3)/2$ denotes the mass per junction, and 
$\phi_{j}$ is the oscillation of a junction $j$.
The action per junction may be smeared on the two-dimensional lattice plane,
and we may introduce a coordinate $(z_1,z_2)$, which is 
just the target space coordinate restricted on the lattice plane, to
parameterize the collective field for the transverse oscillation.
Then \eq{kinact1} will be rewritten in the form
\be
\label{actcol}
K_{col}=\int dtd^2z\ 
\frac{m}{2s} \left(\frac{d\phi_{col}}{dt}\right)^2,
\ee
where $s$ is the area per vertex, and is explicitly given by
\bea
s&=&\frac{\sqrt{q}}{4t_1t_2t_3}(l_1l_2t_3+l_2l_3t_1+l_3l_1t_2), \cr
q&=&(t_1+t_2+t_3)(t_1+t_2-t_3)(t_2+t_3-t_1)(t_3+t_1-t_2).
\eea

On the other hand, after a straightforward calculation,
the determinant of the metric tensor \eq{metricanot} is given by
\be
g^{-1}={\rm Det}(g^{mn})=\frac{(l_1l_2t_3+l_2l_3t_1+l_3l_1t_2)q}
{4t_1t_2t_3(t_1l_1+t_2l_2+t_3l_3)^2}.
\ee
Hence the action for the collective field \eq{actcol} is
\be
\label{keq}
K_{col}=\frac12 \int dtd^2z\ \sqrt{g} 
\left(\frac{t_1t_2t_3}{l_1l_2t_3+l_2l_3t_1+l_3l_1t_2}\right)^{\frac12}
\left(\frac{d\phi_{col}}{dt}\right)^2.
\ee
We do not have any explanations for the extra factor in \eq{keq}. 
Anyway, in our present approximation without any interactions and a coordinate
dependence of the lattice variables $t_i,l_i$, 
we may freely rescale the fields. 
Since the collective field $\phi_{col}$ has mass dimension $-1$, 
it is natural to rescale this field by a factor with a  
non-zero mass dimensions 
and define a new field with the canonical dimension 1/2:
\be
\phi=\left(\frac{t_1t_2t_3}{l_1l_2t_3+l_2l_3t_1+l_3l_1t_2}
\right)^{\frac14}\phi_{col}.
\ee
The potential term can be determined from the condition that the 
energy-momentum relation be generated correctly. Thus the effective action 
of the transverse modes should be
\be
S_{transverse}=-\int d^3x \sqrt{-det(g_{\mu\nu})}\ 
\sum_{i=1}^7 
\left(g^{\mu\nu}\frac{\pt \phi^i}{\pt x^\mu}\frac{\pt \phi^i}{\pt x^\nu}
\right),
\ee
where
\be
g^{\mu\nu}=\left(
\begin{array}{cc}
-1 & 0 \cr
0& g^{mn}
\end{array}
\right)
\ee
and $x^\mu=(t,z^i)$.

As for the tangent modes, the collective field is given by a 
two-dimensional vector field. Following the same discussions given in our 
previous paper \cite{sasa}, the potential term of the low energy Lagrangian
has the gauge symmetry studied in the previous sections, and hence   
the effective theory is expected to be Maxwell theory. 
The action is 
\be
S_{tangent}=-\int d^3x \sqrt{-det(g_{\mu\nu})}\ \frac14
g^{\mu\nu}g^{\rho\sigma}F_{\mu\rho}F_{\nu\sigma}.
\ee
The kinetic term of this action takes the form
\be
\frac12 g_{mn}\frac{\pt a^m}{\pt t}\frac{\pt a^n}{\pt t},
\ee
where $a^m\ (m=1,2)$ 
is the gauge potential, which represents the collective in-plane
motion of the string lattice in the target space-time. 
This kinetic term cannot be derived
from a similar simple argument as above for the scalar fields 
that an average mass $m$ is assigned to each junction.
This fact seems to indicate a subtlety of the in-plane string dynamics, 
and another more careful treatment is necessary.

Since the metric $g^{\mu\nu}$ is just constant in our periodic string lattice, 
from now on, we change the coordinate and work in a coordinate with 
Minkowski metric $g^{\mu\nu}=diag(-1,1,1,1)$ in the following discussions
about the fermionic sector and supersymmetries.
The spectra of the fermionic sector on a string lattice have also been 
discussed in \cite{Callan}.
There they showed that the remaining world-sheet supersymmetries on strings
are consistent with the boundary conditions at junctions, so that there is a 
one-to-one correspondence between the bosonic and fermionic spectra.
Thus there exist one fermionic low energy propagation mode satisfying
$\omega=\pm\sqrt{g^{mn}p_m p_n}$ per each of the transverse directions
and one in-plane fermionic mode.
The simplest fermion in (2+1)-dimensions is a Majorana fermion, which
has two real component. The Dirac equation for a Majorana fermion generates 
just what we want as the spectra for the low energy fermionic modes 
on a string lattice. Thus the action for the fermion should be
\be
S_{fermion}=\int d^3x \sum_{\beta=1}^8 \psi_\beta^T 
\Gamma^0 \Gamma^\mu \pt_\mu \psi_\beta,
\ee
where we may take any two-by-two real representation of $\Gamma^\mu$.

On the other hand, Sen discussed 
the remaining supersymmetries by a supergravity argument \cite{sen}.
The remaining supersymmetries are the solutions of  
\bea
\label{remsusy}
\epsilon_L&=&\Gamma_{(10)}^0\Gamma_{(10)}^1 \epsilon_L, \cr
\epsilon_R&=&-\Gamma_{(10)}^0\Gamma_{(10)}^1 \epsilon_R, \cr
\epsilon_L&=&\Gamma_{(10)}^0\Gamma_{(10)}^2\ \epsilon_R, 
\eea 
where $\Gamma_{(10)}^\mu$ are the (9+1)-dimensional gamma matrices, and 
$\epsilon_L$ and $\epsilon_R$ are the Majorana-Weyl spinor ${\bf 16}$
of $SO(9,1)$.
Studying \eq{remsusy}, there remain eight supersymmetries, and
the solutions to the first equation of \eq{remsusy} can be taken
as the independent components. 
In our case, the ten-dimensional space-time is divided into the 
(2+1)-dimensional lattice space-time and its transverse seven-dimensional 
space.
Since ${\bf 16}$ of $SO(9,1)$ is $\bf{2}\times \bf{8}$ of 
$SO(2,1)\times SO(7)$ and the first equation of \eq{remsusy} gives 
a condition just for the $SO(2,1)$ part\footnote{This $SO(2,1)$ 
symmetry should not be confused with the same symmetry of the effective action,
because the propagation velocity of the low energy modes is different
from the light velocity of the target space.}, the remaining supersymmetries
are in the spinor representation ${\bf 8}$ of $SO(7)$.
On the other hand the gauge field and the scalar fields in the effective
action is in the singlet and vector representation of $SO(7)$, respectively.
To combine the bosonic and fermionic fields into a multiplet 
the fermions in the effective
action should be in the spinor representation.
This pattern of the multiplet is just what we will obtain from the 
dimensional reduction of the (9+1)-dimensional $N=1$ vector multiplet
to (2+1)-dimensions.
Thus we conclude that the low energy effective theory is just 
given by the dimensional reduction of the $N=1$ (9+1)-dimensional super  
Maxwell theory.
 
This result sounds rather curious, since the number of the supersymmetries of
the effective action should be eight rather than sixteen.
In the low-energy effective action there is the two-dimensional rotational
symmetry, and so there are no good ways to impose non-rotationally
symmetric conditions as those in \eq{remsusy} on the sixteen supersymmetries
of the effective action. On the other hand at sufficiently high energy, 
the bare lattice structure will contribute to the dynamics.
Thus it is expected that, although 
the supersymmetries are double in the  effective action within 
our range of approximation of this paper, 
the higher order terms relevant at sufficiently high energy
are expected to break a half of them.

\section{Summary and discussions}
\label{sum}
In this paper, we have discussed the low energy effective geometry on 
a two-dimensional string lattice, 
by studying the energy-momentum relations of the low energy propagation 
modes.
An interesting result is that the geometry is the same for the 
tangent and transverse oscillations, which makes it possible
to assign one geometry to a given string lattice.

We have also discussed the effective action of the low energy propagation
modes. We have obtained the dimensional reduction of $N=1$ 
$(9+1)$-dimensional supersymmetric Maxwell theory. 
A half of the sixteen supersymmetries are expected
to be broken at high energy, where higher order terms out of our range
of approximation become relevant.

We should be careful about the fact that all the results in this paper 
are based on the assumption that the oscillations of a string lattice 
are small.
This does not seem to be a justifiable assumption because the zero modes 
may  considerably change the local structure of the lattice
and even its topology\footnote{This topology change in type IIB string 
theory can be viewed in a different interesting way 
when a string network is lifted to M-theory. 
See \cite{zerodyn} for the recent discussions.}.
As one can see from the explicit expression of the geometry \eq{metricanot}, 
the changes of the string lengths change the low energy effective geometry 
on a string lattice, and so
the zero mode dynamics might be related to the gravity side of 
the effective field theory.
This argument is yet very uncertain, and it seems an important open
matter to identify the roles of these zero modes in the low energy 
effective action.

%%%%%%%%%%%%%%%%%%%%%%%%%%%%%%%%%%%%%%%%%%%%%%%%%%%%%%%%%%%%%
\vspace{.5cm}
\noindent
{\large\bf Acknowledgments}\\[.2cm]
The author is supported in part by the Fellowship Program 
for Japanese Scholars and Researchers to study abroad, 
in part by Grant-in-Aid for Scientific Research
(\#12740150), and in part by Priority Area:
``Supersymmetry and Unified Theory of Elementary Particles'' (\#707),
from Ministry of Education, Science, Sports and Culture, Japan.


\begin{thebibliography}{99}

\bibitem{junction1}
J.~H.~Schwarz,
``Lectures on superstring and M theory dualities,''
Nucl.\ Phys.\ Proc.\ Suppl.\ {\bf 55B}, 1 (1997)
[hep-th/9607201].
%%CITATION = HEP-TH 9607201;%%

\bibitem{junction2}
O.~Aharony, J.~Sonnenschein and S.~Yankielowicz,
``Interactions of strings and D-branes from M theory,''
Nucl.\ Phys.\ {\bf B474}, 309 (1996)
[hep-th/9603009].
%%CITATION = HEP-TH 9603009;%%


\bibitem{BPS1}
K.~Dasgupta and S.~Mukhi,
``BPS nature of 3-string junctions,''
Phys.\ Lett.\ {\bf B423}, 261 (1998)
[hep-th/9711094].
%%CITATION = HEP-TH 9711094;%%

\bibitem{sen}
A.~Sen,
``String network,''
JHEP{\bf 9803}, 005 (1998)
[hep-th/9711130].
%%CITATION = HEP-TH 9711130;%%

\bibitem{zero1}
O.~Aharony, A.~Hanany and B.~Kol,
``Webs of (p,q) 5-branes, five dimensional field theories and grid  diagrams,''
JHEP{\bf 9801}, 002 (1998)
[hep-th/9710116].
%%CITATION = HEP-TH 9710116;%%

\bibitem{zero2}
A.~Mikhailov, N.~Nekrasov and S.~Sethi,
``Geometric realizations of BPS states in N = 2 theories,''
Nucl.\ Phys.\ {\bf B531}, 345 (1998)
[hep-th/9803142].
%%CITATION = HEP-TH 9803142;%%

\bibitem{kol}
B.~Kol,
``Thermal monopoles,''
JHEP{\bf 0007}, 026 (2000)
[hep-th/9812021].
%%CITATION = HEP-TH 9812021;%%

\bibitem{sasa}
N.~Sasakura,
``Low-energy propagation modes on string network,''
hep-th/0012270.
%%CITATION = HEP-TH 0012270;%%

\bibitem{brane}
N.~Arkani-Hamed, S.~Dimopoulos and G.~Dvali,
``The hierarchy problem and new dimensions at a millimeter,''
Phys.\ Lett.\ {\bf B429}, 263 (1998)
[hep-ph/9803315].
%%CITATION = HEP-PH 9803315;%%

\bibitem{braneW}
L.~Randall and R.~Sundrum,
``Out of this world supersymmetry breaking,''
Nucl.\ Phys.\ {\bf B557}, 79 (1999)
[hep-th/9810155].
%%CITATION = HEP-TH 9810155;%%

\bibitem{Regge}
T.~Regge,
``General Relativity Without Coordinates,''
Nuovo Cim.\ {\bf 19} (1961) 558.
%%CITATION = NUCIA,19,558;%%

\bibitem{Penrose}
R. Penrose, in {\it Quantum theory and beyond} ed. T. Bastin,
Cambridge U Press 1971.

\bibitem{gravity1}
C.~Rovelli and L.~Smolin,
``Spin networks and quantum gravity,''
Phys.\ Rev.\ D {\bf 52}, 5743 (1995)
[gr-qc/9505006].
%%CITATION = GR-QC 9505006;%%
 
\bibitem{gravity2}
H.~Ooguri and N.~Sasakura,
``Discrete and continuum approaches to three-dimensional quantum gravity,''
Mod.\ Phys.\ Lett.\ {\bf A6}, 3591 (1991)
[hep-th/9108006].
%%CITATION = HEP-TH 9108006;%%

\bibitem{Ambjorn}
J.~Ambjorn, B.~Durhuus and T.~Jonsson,
``Quantum geometry. A statistical field theory approach,''
{\it  Cambridge, UK: Univ. Pr. (1997) 363 p}.

\bibitem{Rey}
S.~Rey and J.~Yee,
``BPS dynamics of triple (p,q) string junction,''
Nucl.\ Phys.\ {\bf B526}, 229 (1998)
[hep-th/9711202].
%%CITATION = HEP-TH 9711202;%%

\bibitem{Callan}
C.~G.~Callan and L.~Thorlacius,
``Worldsheet dynamics of string junctions,''
Nucl.\ Phys.\ {\bf B534}, 121 (1998)
[hep-th/9803097].
%%CITATION = HEP-TH 9803097;%%

\bibitem{zerodyn}
P.~Shocklee and L.~Thorlacius,
``Zero-mode dynamics of string webs,''
hep-th/0101080.
%%CITATION = HEP-TH 0101080;%%

\end{thebibliography}
\end{document}